\documentclass[prd,preprint,superscriptaddress,showpacs,nofootinbib,%
tightenlines]{revtex4}
\usepackage{epsfig}
\usepackage{slashed}
\usepackage{amssymb}

\begin{document}

\title{Magnetic dipole moment of the $\Delta(1232)$ in chiral
perturbation theory}
\author{C.~Hacker}
\affiliation{Institut f\"ur Kernphysik, Johannes
Gutenberg-Universit\"at, J.~J.~Becher-Weg 45,  D-55099 Mainz,
Germany}
\author{N.~Wies}
\affiliation{Institut f\"ur Kernphysik, Johannes
Gutenberg-Universit\"at, J.~J.~Becher-Weg 45,  D-55099 Mainz,
Germany}
\author{J.~Gegelia}
\affiliation{Institut f\"ur Kernphysik, Johannes
Gutenberg-Universit\"at, J.~J.~Becher-Weg 45,  D-55099 Mainz,
Germany} \affiliation{ High Energy Physics Institute, Tbilisi
State University,
Georgia}
\author{S.~Scherer}
\affiliation{Institut f\"ur Kernphysik, Johannes
Gutenberg-Universit\"at, J.~J.~Becher-Weg 45,  D-55099 Mainz,
Germany}
\date{March 31, 2006}

\begin{abstract}
   The magnetic dipole moment of the $\Delta (1232)$ is calculated in
the framework of manifestly Lorentz-invariant baryon chiral
perturbation theory in combination with the extended on-mass-shell
renormalization scheme.
   As in the case of the nucleon, at leading order both isoscalar and
isovector anomalous magnetic moments are given in terms of two
low-energy constants.
   In contrast to the nucleon case, at next-to-leading order the isoscalar
anomalous magnetic moment receives a (real) loop contribution.
   Moreover, due to the unstable nature of the $\Delta (1232)$,
at next-to-leading order the isovector anomalous magnetic moment
not only receives a real but also an imaginary loop contribution.
\end{abstract}
\maketitle

\section{Introduction}

   The $\Delta(1232)$ resonance is the most prominent and best studied
nucleon resonance.
   It plays an important role in the phenomenological description of
low- and medium-energy processes.
   This is due to the strong coupling of the $\Delta(1232)$ to the $\pi N$
channel and the relatively small mass difference between the
nucleon and the $\Delta(1232)$.
   The strong decay into a nucleon and a pion results in an extremely
short lifetime and makes a precise determination of such a
fundamental physical quantity as the magnetic dipole moment
nontrivial.
   While the magnetic moments of (almost) stable particles may
be determined by means of spin precession measurements, for
unstable particles this is not possible. Here, one has to resort
to indirect measurements making use of a superior physical
reaction into which the electromagnetic interaction of the
particle in question is embedded as a building block.

   The magnetic moment of the $\Delta^{++}(1232)$ has been
investigated experimentally by measuring the $\pi^+ p$
bremsstrahlung reaction \cite{Nefkens:1977eb,Bosshard:1991zp}
which has been analyzed within various theoretical frameworks
\cite{Heller:1986gn,Wittman:1987kb,Lin:1991qk,
LopezCastro:2000cv,LopezCastro:2000ep}.
   The Particle Data Group only makes a rough estimate of the
range the moment is expected to lie within,
$\mu_{\Delta^{++}}=(3.7-7.5)\,\mu_N$
\cite{Eidel},\footnote{$\mu_N$ denotes the nuclear magneton
$e/(2m_p)$.} while SU(6) symmetry predicts for a member of the
decuplet with charge $eQ$ the value $\mu=Q\mu_p$ ($\mu_p$: proton
magnetic moment) \cite{Beg:1964nm}, resulting for the
$\Delta^{++}$ in $\mu_{\Delta^{++}}=5.58\,\mu_N$.
   The magnetic moment of the $\Delta^+(1232)$ is accessed in the
reaction $\gamma p\rightarrow  p \pi^0 \gamma'$ which has been
measured by the A2/TAPS collaboration at MAMI
\cite{Kotulla:2002cg}.
   Using theoretical input based on the
phenomenological model of Ref.\ \cite{Drechsel:2001qu} the
extracted value reads
$\mu_{\Delta^+}=(2.7^{+1.0}_{-1.3}(\text{stat.})\pm
1.5(\text{syst.})\pm 3(\text{theor.}))\,\mu_N$
\cite{Kotulla:2002cg} (see also Refs.\
\cite{Machavariani:1999fr,Drechsel:2000um,Chiang:2004pw,
Machavariani:2005vn,Pascalutsa:2004je} for additional theoretical
approaches to $\gamma p\rightarrow p \pi^0 \gamma'$).

   On the theoretical side, predictions for the delta magnetic moment
have been obtained in various approaches such as SU(6) symmetry
\cite{Beg:1964nm}, several quark models
\cite{Mitra:1984qx,Krivoruchenko:1984xy,Schlumpf:1993rm,Hong:1994nj,%
Sahoo:1995kg,Linde:1995gr,Buchmann:1996bd,Kim:1997ip,Linde:1997ni,%
Berger:2004yi,Yang:2004jr,He:2004kg}, the Skyrme model
\cite{Kim:1989qc}, the $1/N_c$ expansion \cite{Lebed:2004fj},
lattice QCD \cite{Leinweber:1992hy,Cloet:2003jm,Lee:2005ds}, QCD
sum rules \cite{Lee:1997jk,Aliev:2000rc}, heavy-baryon chiral
perturbation theory (HBChPT)
\cite{Butler:1993ej,Banerjee:1995wz,Cloet:2003jm}, quenched ChPT
\cite{Arndt:2003we}, and chiral effective field theory
\cite{Pascalutsa:2004je}.
   The aim of this letter is to calculate the
magnetic moment of the $\Delta(1232)$ up to and including chiral
order $p^3$ in a manifestly Lorentz-invariant formulation of
baryon chiral perturbation theory with explicit $\Delta$ degrees
of freedom ($\Delta$ChPT) \cite{Hacker:2005fh}.\footnote{Here, $p$
stands for small parameters of the theory like the pion mass and
the $\Delta$-nucleon mass difference.}
   Our approach differs from that
of a previous manifestly Lorentz-invariant calculation
\cite{Pascalutsa:2004je} in the structure of the effective
Lagrangian, the power counting scheme, and the renormalization
scheme.
    In Sec.\ II we introduce the relevant effective Lagrangian
and state the power counting.
   In Sec.\ III we calculate the magnetic moment of
the $\Delta(1232)$ at ${\cal O}(p^3)$.
   Section IV contains a short summary.

\section{Effective Lagrangian}
\label{el}
   The effective Lagrangian and the power counting relevant
to classifying the renormalized diagrams for the calculation of
the magnetic moment of the $\Delta(1232)$ have been discussed in
Ref.~\cite{Hacker:2005fh}.
   The non-resonant part of the effective
Lagrangian is that of BChPT with only pion and nucleon fields
\cite{Gasser:1988rb} (see, e.g., Refs.~\cite{Scherer:2002tk} for
an introduction).
   All parameters and fields are considered as
renormalized quantities in the extended on-mass-shell (EOMS)
renormalization scheme of Ref.\ \cite{Fuchs:2003qc}.
   The effective Lagrangian of the $\Delta(1232)$ resonance
[$I(J^P)=\frac{3}{2}(\frac{3}{2}^+)$] is formulated in terms of
vector-spinor isovector-isospinor Rarita-Schwinger fields
$\Psi_{\mu,i}$ \cite{Rarita:1941mf}.
   The most general Lagrangian depends on a free
"off-shell parameter" $A$ \cite{Moldauer:1956}.
   Analyzing the constraints required for obtaining the
correct number of physical degrees of freedom one finds that not
all coupling constants of the original most general Lagrangian are
independent \cite{Wies:2006rv}.
   The relations among the coupling constants also involve the parameter
$A$, however, in such a way that the resulting effective
Lagrangian is invariant under the set of "point transformations"
(see, Refs.~\cite{Hacker:2005fh} and \cite{Wies:2006rv} for
further details).
   As a result of this invariance, physical quantities do not depend on
$A$ and we are free to choose a convenient value for $A$, say,
{$A=-1$}.

   For this choice of $A$, the leading-order Lagrangian reads
\begin{equation}
{{\cal L}_{\Delta}^{(1)}=\bar\Psi_\mu \xi^{\frac{3}{2}}
\Lambda^{\mu\nu}\xi^{\frac{3}{2}} \Psi_\nu,} \label{deltaL}
\end{equation}
with the isospin projection operator $\xi^{\frac{3}{2}}_{ij} =
\delta_{ij} -\frac{1}{3}\tau_i \tau_j$ and
\begin{eqnarray}
\label{Ldeltafree} \Lambda_{\mu\nu} &=& -\Biggl\{ \left( i
D\hspace{-.70em}/\hspace{.1em}-m_{\Delta}\right)\,g_{\mu\nu} -i
\,\left( \gamma_{\mu}D_{\nu}+\gamma_{\nu}D_{\mu}\right)\nonumber \\
&&+i\,\gamma_{\mu} D\hspace{-.70em}/\hspace{.1em}\gamma_{\nu} +
m_{\Delta}\,\gamma_{\mu}\gamma_{\nu}+\frac{g_1}{2}\,\biggl[
u\hspace{-.5em}/\, g_{\mu\nu}-\gamma_\mu u_\nu - u_\mu \gamma_\nu
+\gamma_\mu u\hspace{-.5em}/\,\gamma_\nu\biggr]\,\gamma_5
\Biggr\}.
\end{eqnarray}
   The covariant derivative of the delta field is defined as
\begin{eqnarray*}
(D_\mu \Psi)_{\nu,i}&\equiv&
\partial_\mu\Psi_{\nu,i}
-2i\epsilon_{ijk}\Gamma_{\mu,k}\Psi_{\nu,j}
+\Gamma_{\mu}\Psi_{\nu,i} -iv_\mu^{(s)}\Psi_{\nu,i}
\end{eqnarray*}
and involves the connection $\Gamma_\mu=\frac{1}{2}\, \left[
u^{\dagger}\left(\partial_\mu-i r_\mu\right)u + u \left(
\partial_\mu-i l_\mu\right) u^{\dagger}\right]=\tau_k\Gamma_{\mu,k}$.
   The pion fields are contained in the unimodular unitary $(2\times 2)$
matrix $U$ with $u^2=U$.
   In case of the electromagnetic interaction, we insert for the
external fields $r_\mu=l_\mu=-e\frac{\tau_3}{2}{\cal A}_\mu$ and
$v_\mu^{(s)}=-\frac{e}{2}{\cal A}_\mu$ \cite{Scherer:2002tk},
where $e$ is the proton charge.
   The $\pi\Delta\Delta$ interaction is generated by
the last term of Eq.\ (\ref{Ldeltafree}), where $u_\mu=i\left[
u^{\dagger}\left(\partial_\mu-i r_\mu\right)u-u \left(
\partial_\mu-i l_\mu \right) u^{\dagger}\right]=\tau_k u_{\mu,k}$
and $g_1$ is the relevant coupling constant.
   Finally, {$m_\Delta $} stands for the mass of the $\Delta$.
   The leading-order {$\pi N\Delta$} interaction Lagrangian reads
\begin{equation}
{{\cal L}_{\pi N\Delta}^{(1)}= - g \,\bar{\Psi}_{\mu,i}
\,\xi^{\frac{3}{2}}_{ij} \,(g^{\mu\nu}
-\gamma^{\mu}\gamma^{\nu})\, u_{\nu,j}\, \Psi +h.c.\,,}
\label{pND}
\end{equation}
where $\Psi=(p,n)^T$ denotes the nucleon field with two
four-component Dirac fields $p$ and $n$ describing the proton and
neutron, respectively, and {$g$} is a coupling constant.
   Finally, for the calculations of this work it is sufficient to
parameterize the photon-delta interaction Lagrangian at ${\cal
O}(p^2)$ as
\begin{equation}
{\cal
L}_{\Delta}^{(2)}=\frac{i\,e}{2\,m_\Delta}\,\bar\Psi_\mu\xi^{\frac{3}{2}}
\left[1+\frac{3}{2}\,d_1+ 3\left(1+\frac{3}{2}\,
d_2\right)\tau_3\right]\xi^{\frac{3}{2}} \Psi_\nu\,{\cal
F}^{\mu\nu}\,, \label{phdlagr}
\end{equation}
where ${\cal F}^{\mu\nu}$ denotes the field-strength tensor and
$d_1$, $d_2$ are coupling constants contributing to the magnetic
dipole moment of the $\Delta$ at the given order.\footnote{The
separation has been introduced for later convenience so that the
final expression of the magnetic moment in the usual isospin basis
is most simple.}

  The perturbative calculation of the dipole moment is organized by
applying the following power counting to the renormalized
diagrams.
  Interaction vertices obtained from an ${\cal O}(p^n)$ Lagrangian
count as order $p^n$, a pion propagator as order $p^{-2}$, a
nucleon propagator as order $p^{-1}$, and the integration of a
loop as order $p^4$.
   In addition, we assign the order $p^{-1}$ to the
$\Delta$ propagator and the order $p^1$ to the mass difference
$\delta\equiv m_\Delta-m$.

   In a resonance generating channel, a $\Delta$ propagator which
is not not involved in a loop integration has to be dressed.
   One then has to re-sum the self-energy insertions and to
   consider the dressed propagator as of the order $p^{-3}$,
because the self-energy starts at ${\cal O}(p^3)$.

\section{Magnetic moment of the $\Delta(1232)$}

   Unstable particles do not occur in the spectrum of asymptotic
states of the theory.
   Therefore the standard definition of the
magnetic moment through the matrix element of the current between
asymptotic free states cannot be applied.
   Instead one considers a complete physical scattering amplitude
where the unstable particle contributes as an intermediate state.
   One parameterizes the contribution of the unstable particle and
defines the magnetic moment such that for the regime where the
unstable particle turns into a stable one (here
$m_\Delta<m_N+M_\pi$) the magnetic moment coincides with the
standard one of the stable particle.

\begin{figure}[t]
\vspace{0mm} \hspace*{1. cm} \epsfig{file=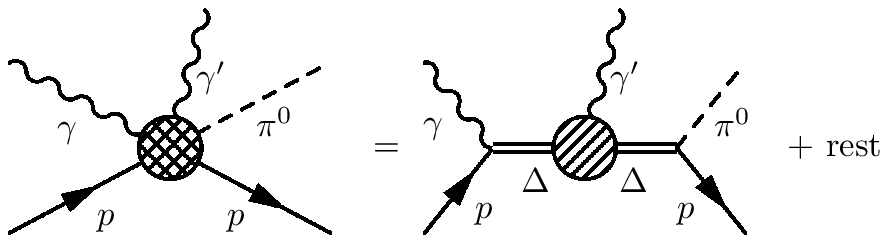,width=9 cm}
\caption[]{\label{vertex:fig} $\gamma p\to p \pi^0 \gamma'$
amplitude in the $\Delta(1232)$ resonance
region.\label{diagram.fig}}
\end{figure}

   For example, for the $\Delta^+(1232)$ resonance one considers the
physical process
\begin{equation}
\label{process} {\gamma + p\to p +\pi^0+\gamma'}\,.
\end{equation}
   In the delta resonance region, to leading orders ($p^{-3}$,
$p^{-2}$, $p^{-1}$) in $\Delta$ChPT the contribution of the
$\Delta^+(1232)$ can be consistently separated.
   The contribution shown in Fig.~\ref{diagram.fig}
is symbolically of the form
\begin{eqnarray}
&&V_\delta\, S^{\delta\gamma}\,\Gamma^\nu_{\gamma\beta}\,
S^{\beta\alpha} V_\alpha^\mu\,, \ \  \label{deltacontr}
\end{eqnarray}
where $V^\mu$, $\Gamma^\nu$, and $V$ denote the $\gamma N\Delta$,
$\gamma\Delta\Delta$, and $\pi\Delta N$ vertices, respectively,
and $S$ corresponds to the $\Delta$ propagator.
   We parameterize the (most general) $\gamma\Delta\Delta$ vertex
$\Gamma^\nu$ in terms of the Lorentz structures
$$
\Gamma^\nu_{\gamma\beta}=g_{\gamma\beta} \Gamma^\nu_1(p',p) +
\gamma_\gamma \Gamma^\nu_2(p',p)\gamma_\beta+ p'_\gamma
\Gamma^\nu_3(p',p) p_\beta +\cdots\,
$$
and expand around {$p^2=m_\Delta^2,p
\hspace{-.45em}/\hspace{.1em}=m_\Delta,p'^2=m_\Delta^2,p
\hspace{-.45em}/\hspace{.1em}'=m_\Delta$}.
   Then, the only relevant contribution up to and including
the next-to-next-to-leading order reads\footnote{Note that
$\gamma^\alpha S_{\alpha\beta}(p)$ and $p^\alpha
S_{\alpha\beta}(p)$ are free of poles and therefore generate only
terms of higher order.}
\begin{equation}
{ V_\delta\, S^{\delta\gamma}\,g_{\gamma\beta}\,\Gamma^\nu_1
(p',p)\Big|_{p^2=m_\Delta^2,p
\hspace{-.35em}/\hspace{.1em}=m_\Delta,p'^2=m_\Delta^2,p'
\hspace{-.55em}/\hspace{.1em}=m_\Delta}\, S^{\beta\alpha}
V_\alpha^\mu\, .} \label{Vparametrization}
\end{equation}
   As a result of Eq.~(\ref{Vparametrization}), at leading
orders ($p^{-3}$, $p^{-2}$, $p^{-1}$) one can consider the
$\gamma\Delta\Delta$ vertex function with an ''on-mass-shell
{$\Delta$}'' and parameterize
\begin{eqnarray}
\label{DFFDef} {\Gamma_1^\nu (p',p)} & = & \gamma^{\nu}F(Q^2)
+\frac{(p+p')^\nu}{2m_\Delta} \ G(Q^2)\,+\cdots,\nonumber\\
&& q^\nu=(p'-p)^\nu,\quad Q^2=-q^2\,,
\end{eqnarray}
where the complete on-shell vertex contains two additional
structures \cite{Nozawa:1990gt} which are, however, not related to
the magnetic moment.
   We express the total magnetic moment as
\begin{equation}
\label{vecmu} \vec\mu=[Q+\kappa]\, \frac{e}{2m_\Delta}\, g\,
\vec{S},
\end{equation}
where $\vec S$ is the spin, $eQ$ is the charge, and $g=2$ (in
combination with $\kappa=0$) is the gyromagnetic ratio of a
particle that does not participate in the strong interactions,
neglecting also higher order weak and electromagnetic interactions
\cite{Weinberg:1970}.\footnote{The use of minimal substitution
only generates $g=2/3$ instead of $g=2$.}
   Here, we will only consider the modification due to the strong
interactions which are encoded in the anomalous magnetic moment
$\kappa$.
   The total magnetic moment in units of $e/(2 m_\Delta)$
is given by $F(0)$ of Eq.\ (\ref{DFFDef}).
   Performing an isospin
decomposition in the isovector-isospinor representation as
\begin{displaymath}
F(Q^2)=\frac{1}{2}F^{(s)}(Q^2)+\frac{3}{2}\tau_3 F^{(v)}(Q^2),
\end{displaymath}
one obtains for the isoscalar and isovector components of the
magnetic dipole moment
\begin{equation}
\mu_\Delta^{(s)}=F^{(s)}(0) \,\frac{e}{2 m_\Delta}, \ \
\mu_\Delta^{(v)}=F^{(v)}(0)\,\frac{e}{2 m_\Delta}\,.
\label{magneticmomenta}
\end{equation}
   The magnetic dipole moment of the physical degrees of freedom is given by
\begin{equation}
\mu=\frac{1}{2}\,\mu_\Delta^{(s)}+T_3\,
\mu_\Delta^{(v)}=3\left[\frac{1}{2}\,\left(
1+\kappa^{(s)}_\Delta\right)+T_3\, \left(
1+\kappa^{(v)}_\Delta\right)\right]\, \frac{e}{2 m_\Delta}\,,
\label{mm}
\end{equation}
where $T_3$ stands for the third component of the isospin operator
in the usual four-dimensional representation.

\begin{figure}
\begin{center}
\epsfig{file=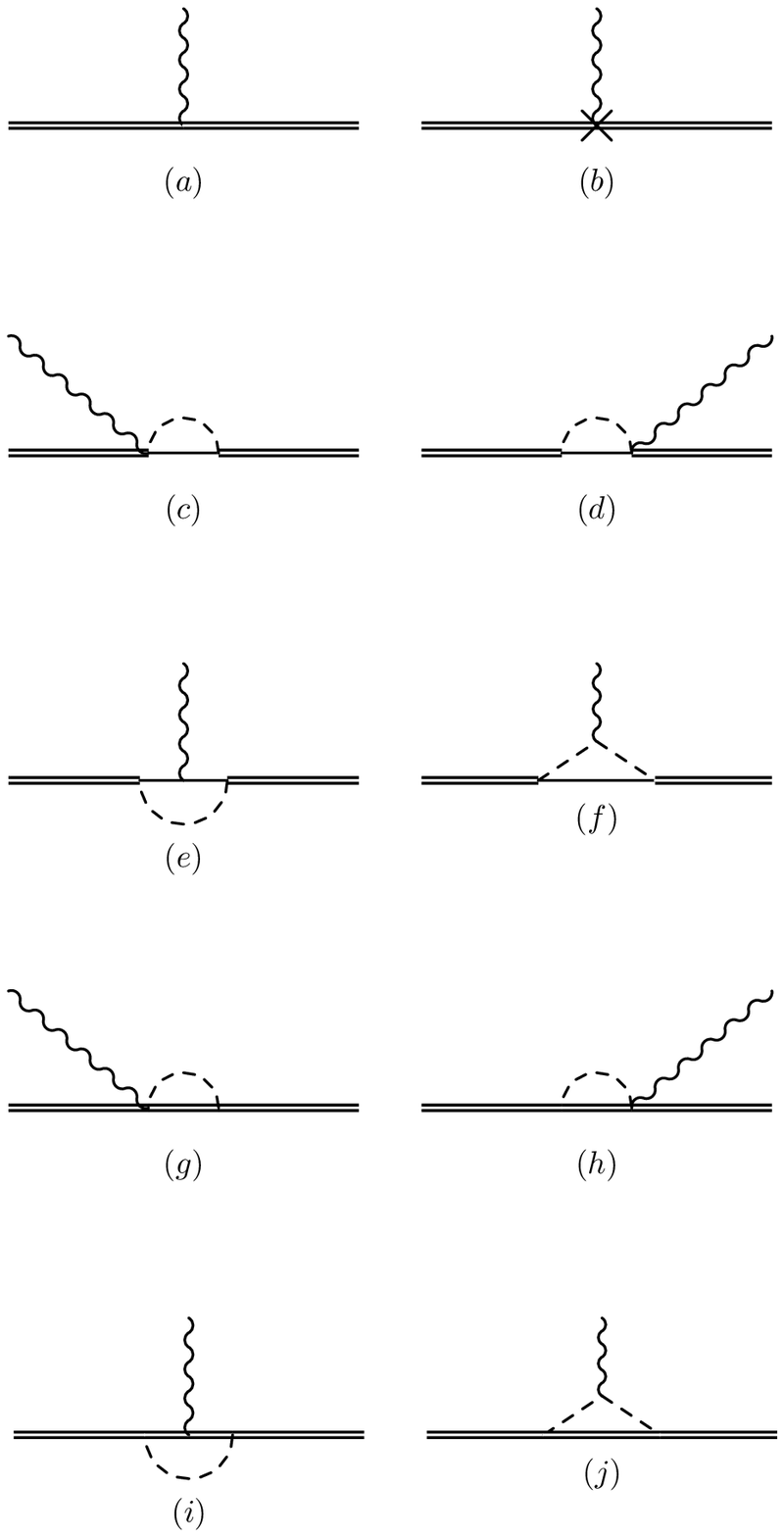}
\end{center}
\caption[]{\label{diagrams} Contributions to the
$\gamma\Delta\Delta$ vertex up to and including
$\mathcal{O}(p^3)$.}
\end{figure}

   Using the Lagrangians of Sec.\ \ref{el}, we have calculated the
$\gamma \Delta\Delta$ vertex up to and including
$\mathcal{O}(p^3)$, where the relevant diagrams are shown in Fig.\
\ref{diagrams}.
   Applying the EOMS renormalization scheme
\cite{Fuchs:2003qc},\footnote{The contribution of diagram
\ref{diagrams} (i) to the magnetic moment is of higher order.} we
obtain the following renormalized expressions for the form factors
$F^{(s)}$ and $F^{(v)}$ at $Q^2=0$:
\begin{eqnarray}
F^{(s)}(0) &=& 3 + 3 d_1+\frac{71 g^2 m_\Delta \delta}{512\, \pi^2
\,F^2
}+\mathcal{O}(p^4),\nonumber\\
F^{(v)}(0) &=& 3 + 3 d_2 -\frac{g_1^2 M m_\Delta}{54\, \pi\, F^2 }
-\frac{g^2 m_\Delta}{4608\, \pi^2\, F^2}\left[443\,
\delta+384\,\delta\, \ln\left(
\frac{M}{m_\Delta}\right)\right.\nonumber\\
&&\left.+384 \Theta \,\ln\left(
\frac{\delta+\Theta}{M}\right)\right]+\frac{i \, g^2 m_\Delta
\Theta}{12\, \pi \,F^2}+\mathcal{O}(p^4)\,, \label{fsfv}
\end{eqnarray}
where \mbox{$\Theta = \sqrt{-M^2+\delta^2}$}.
   For the numerical analysis we make use of
$g_A=1.267$, $F_\pi=92.4\,\mbox{MeV}$,
$m_N=m_p=938.3\,\mbox{MeV}$, $M_\pi=M_{\pi^+}=139.6\,\mbox{MeV}$,
and $m_\Delta=1210\,\mbox{MeV}$, where $m_\Delta$ is the pole
mass.
   For the delta coupling constants we take $g=1.127$ as
obtained from a fit to the $\Delta\to\pi N$ decay width
\cite{Hacker:2005fh} and $g_1=9g_A/5$ from SU(6) symmetry.
   We then obtain for $\kappa^{(s)}$ and
$\kappa^{(v)}$
\begin{eqnarray}\label{magnetergebnis1}
\kappa^{(s)}_\Delta&=&{d}_1+0.23+\mathcal{O}(p^4),\nonumber \\
\kappa^{(v)}_\Delta&=&{d}_2-0.22+i\,0.37+\mathcal{O}(p^4).
\end{eqnarray}
   Unfortunately we do not have an estimate for the
parameters ${d}_1$ and ${d}_2$ which reflect the contribution to
the anomalous magnetic moment of the $\Delta (1232)$ at tree-level
[${\cal O}(p^2)$].
   However, we can compare the results of Eq.\ (\ref{magnetergebnis1})
with the anomalous magnetic moment of the nucleon
\cite{Fuchs:2003ir}
\begin{eqnarray}\label{magnetergebnis2}
\kappa^{(s)}_N&=&2 c_7 m +{\cal O}(p^4),\nonumber \\
\kappa^{(v)}_N&=&4 c_6 m - \frac{g_A^2 m M}{4\pi F^2}=4 c_6
m-1.96+\mathcal{O}(p^4),
\end{eqnarray}
   where $c_6$ and $c_7$ are parameters of the ${\cal O}(p^2)$
$\pi N$ Lagrangian.
   There are two main differences between the
anomalous magnetic moments of the nucleon and of the $\Delta
(1232)$ up to the chiral order $p^3$.
   First, pion loops do not contribute to $\kappa^{(s)}_N$ at
${\cal O}(p^3)$, whereas the loop contribution to
$\kappa^{(s)}_\Delta$ is 0.23.
   The loop contribution originates from the renormalized
   diagram (e) of Fig.~\ref{diagrams}.
   Second, due to the unstable nature of the $\Delta
(1232)$, there is an imaginary part in $\kappa^{(v)}_\Delta$ which
the nucleon does not have.
   The isovector loop contribution is significantly smaller than
in the case of the nucleon.

   Let us finally compare the results with previous EFT calculations.
   As in the nucleon case, ChPT in the SU(2) sector does not {\em predict}
the anomalous magnetic moments of the $\Delta(1232)$.
   In Refs.\ \cite{Butler:1993ej,Banerjee:1995wz,Cloet:2003jm}
the decuplet magnetic moments were calculated in the heavy-baryon
framework.
   Our calculation differs by the set of diagrams which contibute
at the given order.
   The manifestly Lorentz-invariant approach of
Ref.\ \cite{Pascalutsa:2004je} uses a different power counting and
considers a different set of diagrams.
   The loop contributions obtained in Ref.\ \cite{Pascalutsa:2004je}
turn out to be larger in magnitude than our results.
   Note, however, that different renormalization schemes result in
different values for the low-energy constants.

\section{Summary}

   We have calculated the magnetic dipole moment of the $\Delta
(1232)$ up to and including order $p^3$ treating both the pion
mass and the delta-nucleon mass difference as small quantities of
order $p$.
   For this purpose we have used the manifestly Lorentz-invariant form
of BChPT with explicit $\Delta$ degrees of freedom
\cite{Hacker:2005fh} in combination with the EOMS renormalization
scheme \cite{Fuchs:2003qc}.
   This results in a consistent effective field theory describing
the correct number of physical degrees of freedom in combination
with a systematic power counting.
   The $\pi\Delta\Delta$ interaction was chosen to be consistent
with a recent analysis of the structure of constraints of Ref.\
\cite{Wies:2006rv} for a spin-3/2 system.

   At next-to-leading order, ${\cal O}(p^2)$, the isoscalar and isovector
anomalous magnetic moments are given in terms of two low-energy
constants.
   At next-to-next-to-leading order the isoscalar
anomalous magnetic moment receives a real loop contribution of
$0.18$ in units of the nuclear magneton.
   This has to be contrasted with the nucleon, where the loop
contribution to the isoscalar anomalous magnetic moment is ${\cal
O}(p^4)$.
   At next-to-next-to-leading order the isovector
anomalous magnetic moment receives a real loop contribution of
$-0.17$ and an imaginary loop contribution of $0.29$ in units of
$\mu_N$.
   The appearence of an imaginary part in the $\gamma
\Delta\Delta$ vertex function reflects the unstable nature of the
$\Delta (1232)$.

   As a next step it would be desirable to have full and
consistent calculations of $\pi^+ p$ bremsstrahlung and $\gamma
p\rightarrow  p \pi^0 \gamma'$ in the delta resonance region.
   Such calculations would have the potential of allowing for an
extraction of the parameters $d_1$ and $d_2$ from a fit to the
experimental cross sections and thus for obtaining a result for
the magnetic moments in a self-consistent framework.

\acknowledgments

   We would like to thank Matthias R.~Schindler for useful comments
on the manuscript.
   The work of J.~G.~has been supported by the
Deutsche Forschungsgemeinschaft (contract SCHE 459/2-1).

\end{document}